\documentclass[12pt,a4paper]{article}
\usepackage{amsmath}
\begin{document}

\begin{center}
{\bf Thermal Bogoliubov transformation in nuclear structure theory}
\vskip 5mm A. I.~Vdovin, Alan A. Dzhioev
\vskip 5mm
{\it
Bogoliubov Laboratory of Theoretical Physics, Joint
Institute for Nuclear Research, 141980 Dubna, Russia }
\end{center}
\vskip 5mm \centerline{\bf Abstract } Thermal Bogoliubov
transformation is an essential ingredient of the thermo field
dynamics -- the real time formalism in quantum field and many-body
theories at finite temperatures developed by H. Umezawa and
coworkers. The approach to study properties of hot nuclei which is
based on the extension of the well-known Quasiparticle-Phonon Model
to finite temperatures employing the TFD formalism is presented. A
distinctive feature of the QPM-TFD combination is a possibility to
go beyond the standard approximations like the thermal Hartree-Fock
or the thermal RPA ones. \vskip 10mm


Among numerous outstanding achievements by N.\,N.\,Bogoliubov  there
is the well-known Bogoliubov transformation for bosons
\cite{Bogoliubov_tb} and fermions \cite{Bogoliubov_tf}. This unitary
transformation played a crucial role in constructing microscopical
theories of superfluidity and superconductivity and till now has
been an extremely useful and powerful tool in many branches of
theoretical physics. Quite unexpectedly, a new version of the
Bogoliubov transformation appeared in the middle of the 1970s when
H. Umezawa and coworkers formulated the basic ideas of thermo field
dynamics (TFD) \cite{TFD1, TFD2} -- a new formalism extending the
quantum field and many-body theories to finite temperatures.

%
%

Within TFD \cite{TFD1,TFD2}, the thermal average of a given operator
$A$ is calculated as the expectation value in a specially
constructed, temperature-dependent state $|0(T)\rangle$ which is
termed the thermal vacuum. This expectation value is equal to the
usual grand canonical average of~$A$. In this sense, the thermal
vacuum describes the thermal equilibrium of the system. To construct
the state $|0(T)\rangle$, a formal doubling of the system degrees of
freedom is introduced\footnote{It is worth mentioning the general
statement \cite{Haag67} that the effect of finite temperature can be
included in a free field theory if one doubles the field degrees of
freedom.}. In TFD, a tilde conjugate operator~$\widetilde A$---
acting in the independent Hilbert space --- is associated with $A$,
in accordance with properly formulated tilde conjugation
rules~\cite{TFD1,TFD2,Ojima81}. For a heated system governed by the
Hamiltonian~$H$ the whole Hilbert space is spanned by the direct
product  of the eigenstates of~$H$  and those of the tilde
Hamiltonian~$\widetilde H$, both corresponding to the same
eigenvalues, i.e. ${H|n\rangle=E_n|n\rangle}$ and ${\widetilde
H|\widetilde n\rangle=E_n|\widetilde n\rangle}$. In the doubled
Hilbert space, the thermal vacuum is defined as the zero-energy
eigenstate of the so-called thermal Hamiltonian ${{\mathcal
H}=H-\widetilde H}$. Moreover,  the thermal vacuum satisfies the
thermal state condition~\cite{TFD1,TFD2,Ojima81}
\begin{equation}\label{TSC}
A|0(T)\rangle = \sigma\,{\rm e}^{{\mathcal H}/2T} {\widetilde
A}^\dag|0(T)\rangle,
\end{equation}
where  $\sigma=1$ for bosonic~$A$ and $\sigma=-i$ for fermionic $A$.
It is seen from~(\ref{TSC}) that, in TFD, there always exists a
certain combination of $A$ and $\widetilde A^\dag$ which annihilates
the thermal vacuum. That mixing is promoted by a specific
transformation called the thermal Bogoliubov
transformation~\cite{TFD2}. This transformation must be canonical in
the sense that the algebra of the original system remains the same,
keeping its dynamic. The temperature dependence comes from the
transformation parameters.

The important point is that in the doubled Hilbert space the
time-translation operator is the thermal Hamiltonian~${\mathcal H}$.
This means that the excitations of the thermal system are obtained
by the diagonalization of~${\cal H}$. The existence of the thermal
vacuum annihilation operators provides us with a powerful method to
analyze physical systems at finite temperatures and allows for
straightforward extensions of different zero-temperature
approximations.

In the present note, we exemplify advantages of TFD while treating
the behavior of atomic nuclei at finite temperatures. In particular,
we will show a way of going beyond the thermal RPA and allowing one
to treat a coupling of the basic nuclear modes, quasiparticles and
phonons \cite{Soloviev92}, at finite temperatures. This problem was
already studied  in Refs.~\cite{Tan88,KosVdo94a,KVW97}. However,
 the new aspects have been revealed recently \cite{IJMPE09}.


 To avoid  unnecessary complications in the formulae, we consider
a nuclear Hamiltonian which is a simplified version of the
Hamiltonian of the Quasiparticle-Phonon Model \cite{Soloviev92}. It
consists of a mean field $H_{\rm sp}$, the BCS pairing interaction
$H_{\rm pair}$, and a separable multipole-multipole particle-hole
interaction $H_{\rm ph}$. Moreover, protons and neutrons are not
distinguished. The Hamiltonian reads
\begin{multline}\label{QPM_zero}
 H = H_{\rm sp}+H_{\rm pair}+H_{\rm ph} = \\
 \sum_{jm}(E_j-\lambda)a^\dag_{jm}a^{\phantom{\dag}}_{jm} -\frac14 G\sum_{j_1m_1\,j_2m_2}
 a^\dag_{j_1m_1}a^\dag_{\overline{\jmath_1m_1}}
 a^{\phantom{\dag}}_{\overline{\jmath_2m_2}}a^{\phantom{\dag}}_{j_2m_2}
 - \frac{1}{2}\sum_{\lambda\mu}\kappa_0^{(\lambda)} M^\dag_{\lambda\mu}
M^{\phantom{\dag}}_{\lambda\mu}
 \end{multline}
where $a^\dag_{jm}$ and $a_{jm}$ are the nucleon creation and
annihilation operators, $a_{\overline{\jmath m}}=(-1)^{j-m}a_{
j-m}$, and $M^\dag_{\lambda\mu}$ is the multipole single-particle
operator of the electric type with multipolarity $\lambda$.


At first, we apply TFD to treat pairing correlations at finite
temperature (see also \cite{Civi93,KosVdo94b}). To this aim, we make
the standard Bogoliubov $u, v$-transformation from nucleon operators
to quasiparticle operators $\alpha^\dag, \alpha$
\begin{align}\label{B_tr}
 \alpha^\dag_{jm}&=u_{j}a^\dag_{jm}-v_{j}a_{\overline{\jmath m}},\nonumber\\
 \alpha^{\phantom\dag}_{jm}&=u_{j}a^{\phantom\dag}_{jm}-v_{j}{a}^\dag_{\overline{\jmath
 m}}~~(u^2_j+v^2_j=1)~.
\end{align}
The same transformation with the same $u, v$ coefficients is applied
to nucleonic tilde operators $\widetilde a^\dag_{jm},~\widetilde
a^{\phantom{\dag}}_{jm}$, thus producing the tilde quasiparticle
operators $\widetilde\alpha^\dag_{jm}$ and
$\widetilde\alpha^{\phantom{\dag}}_{jm}$.

Thermal effects appear after the thermal Bogoliubov transformation
which mixes ordinary and tilde quasiparticle operators and produces
the operators of so-called thermal quasiparticles $\beta^\dag_{jm},
\beta_{jm}$ and their tilde counterparts. Following the Ojima's
formulation of the double tilde conjugation rule for fermions
($\widetilde{\widetilde{a}} = a$) \cite{Ojima81} we use here the
complex form of the thermal Bogoliubov transformation:
\begin{align}\label{T_tr}
  \beta^\dag_{jm}&=x_j\alpha^\dag_{jm}- i y_j\widetilde\alpha_{jm}, \nonumber\\
  \widetilde\beta^\dag_{jm}&=x_j\widetilde\alpha^\dag_{jm}+ i y_j\alpha_{jm}~~(x^2_j+y^2_j=1).
\end{align}
The reasons for this are given in~\cite{IJMPE09}.

Then we express the thermal Hamiltonian in terms of thermal
quasiparticle operators (\ref{T_tr}) and require that the one-body
part of the thermal BCS Hamiltonian $\mathcal{H_{\rm BCS}} = H_{\rm
sp} + H_{\rm pair} - \widetilde{H}_{\rm sp} - \widetilde{H}_{\rm
pair}$ becomes diagonal in terms of thermal quasiparticles. This
yields the following expressions for  $u_j,\ v_j$:
\begin{equation}\label{uv}
  u^2_j=\frac{1}{2}\left(1+\frac{E_j-\lambda}%
   {\varepsilon_j}\right),\quad
   v^2_j=\frac{1}{2}\left(1-\frac{E_j-\lambda}%
   {\varepsilon_j}\right),
\end{equation}
where $\varepsilon_j=\sqrt{(E_j-\lambda)^2+\Delta^2}$. The gap
parameter $\Delta$ and the chemical potential $\lambda$ are the
solutions of the equations
\begin{equation}\label{BCS}
\Delta=\frac{G}{2} \sum_j (2j+1)(x^2_j-y^2_j)u_jv_j,\quad N= \sum_j
(2j+1)(v^2_jx^2_j+u^2_jy^2_j),
\end{equation}
where $N$ is the number of nucleons in a nucleus.

With $u_j, v_j$ from (\ref{uv}) the one-body part of the thermal BCS
Hamiltonian reads
\begin{equation}\label{Ht_sp}
{\mathcal H}_{\rm BCS} \simeq \sum_{jm} \varepsilon_j
(\beta^\dag_{jm}\beta^{\phantom{\dag}}_{jm}-\widetilde\beta^\dag_{jm}\widetilde\beta^{\phantom{\dag}}_{jm}).
\end{equation}
One can see that the Hamiltonian $\mathcal H_{\rm BCS}$ describes a
system of noninteracting thermal quasiparticles and
tilde-quasiparticles with energies $\varepsilon_j$ and
$-\varepsilon_j$, respectively.

To determine the thermal vacuum corresponding to $\mathcal H_{\rm
BCS}$, we need to fix appropriately the coefficients $x_j,\ y_j$. In
Refs.~\cite{Civi93,KosVdo94b}, the coefficients were found by
minimizing the thermodynamic potential of the system of
noninteracting Bogoliubov quasiparticles. Here we demand that the
vacuum $|0(T);\mathrm{qp}\rangle$ of thermal quasiparticles obey the
thermal state condition~(\ref{TSC})
\begin{equation}\label{TSC_BSC}
a_{jm}|0(T);\mathrm{qp}\rangle = -i\,{\rm e}^{{\mathcal
H_{\mathrm{BCS}}}/2T}\, {\widetilde
a}^\dag_{jm}|0(T);\mathrm{qp}\rangle.
\end{equation}
Combining~(\ref{TSC_BSC}) and~(\ref{B_tr}) one gets
\begin{eqnarray}\label{occup}
y_j=\left[1+\exp\left(\frac{\varepsilon_j}{T}\right)\right]^{-1/2},
  \quad x_j=\bigl(1-y^2_j\bigr)^{1/2}~.
\end{eqnarray}
We see that the coefficients $y^2_j$ are the thermal Fermi-Dirac
occupation factors which determine the average number of thermally
excited Bogoliubov quasiparticles  in the BCS thermal vacuum.
Equations $(\ref{uv})$, (\ref{BCS}),  and $(\ref{occup})$ are the
well-known finite-temperature BCS equations \cite{Good81}.


In the next stage we partially take into account the particle-hole
residual interaction $H_{\rm ph}$. Now the thermal Hamiltonian reads
\begin{equation}\label{H_full}
{\mathcal H} = \sum_{jm}\varepsilon_j
 (\beta^\dag_{jm}\beta^{\phantom{\dag}}_{jm}-
 \widetilde\beta^\dag_{jm}\widetilde\beta^{\phantom{\dag}}_{jm}) -
 \frac{1}{2}\sum_{\lambda\mu} \kappa^{(\lambda)}_0 \left\{M^\dag_{\lambda\mu}M^{\phantom{\dag}}_{\lambda\mu}
 -\widetilde M^\dag_{\lambda\mu}\widetilde
 M^{\phantom{\dag}}_{\lambda\mu}\right\}
\end{equation}
and it can  be divided into two parts -- ${\mathcal
H}_{\mathrm{TQRPA}}$ and ${\mathcal H}_{\mathrm{qph}}$. The part
${\mathcal H}_{\mathrm{TQRPA}}$ that contains $\mathcal
H_{\mathrm{BCS}}$ and the terms with even numbers of creation and
annihilation operators of thermal quasiparticles is approximately
diagonalized within the Thermal Quasiparticle Random Phase
Approximation, whereas the part ${\mathcal H}_{\mathrm{qph}}$
containing odd numbers of creation and annihilation operators is
responsible for the coupling of TQRPA eigenvectors (thermal
phonons).

 To diagonalize ${\mathcal
H}_{\mathrm{TQRPA}}$, the following operator of thermal phonon is
introduced:
\begin{multline*}
  Q^\dag_{\lambda\mu i}\!=\!\frac12 \sum_{j_1j_2}
 \Bigl(\psi^{\lambda i}_{j_1j_2}[\beta^\dag_{j_1}\beta^\dag_{j_2}]^\lambda_\mu\!\!+
 \widetilde\psi^{\lambda i}_{j_1j_2}[\widetilde\beta^\dag_{\overline{\jmath_1}}
 \widetilde\beta^\dag_{\overline{\jmath_2}}]^\lambda_\mu\!\!+
 2i\eta^{\lambda i}_{j_1j_2}[\beta^\dag_{j_1}
  \widetilde\beta^\dag_{\overline{\jmath_2}}]^\lambda_\mu\Bigr)\\
+(-1)^{\lambda-\mu}\!\left(
 \phi^{\lambda i}_{j_1j_2}[\beta_{j_1}\beta_{j_2}]^\lambda_{-\mu}\!\!+
 \widetilde\phi^{\lambda i}_{j_1j_2}[\widetilde\beta_{\overline{\jmath_1}}
 \widetilde\beta_{\overline{\jmath_2}}]^\lambda_{-\mu}\!\!-
 2i\xi^{\lambda i}_{j_1j_2}[\beta_{j_1}
  \widetilde\beta_{\overline{\jmath_2}}]^\lambda_{-\mu}\!\right),
\end{multline*}
where the notation $[~~]^\lambda_\mu$ means the coupling of
single-particle momenta $j_1$, $j_2$ to the angular momentum
$\lambda$ with the projection $\mu$. Now the thermal equilibrium
state is treated as a vacuum $|0(T);\mathrm{ph}\rangle$ for thermal
phonons. In addition, the thermal phonon operators are assumed to
obey bosonic commutation rules. This imposes some constraints on the
phonon amplitudes $\psi, \widetilde{\phi}, \eta$ etc. (see
Ref.~\cite{IJMPE09} for more details).

To find eigenvalues of ${\mathcal H}_{\mathrm{TQRPA}}$, the
variational principle  is applied, i.e. we find the minimum of the
expectation value of ${\mathcal H}_{\mathrm{TQRPA}}$ with respect to
one-phonon states ${Q^\dag_{\lambda\mu i}|0(T);{\rm ph}\rangle}$ or
${\widetilde Q^\dag_{\overline{\lambda\mu i}}|0(T);{\rm ph}\rangle}$
under afore-mentioned constraints on the phonon amplitudes. As a
result we arrive at the following equation for thermal phonon
energies $\omega_{\lambda i}$:
\begin{gather}\label{TRPA_eq}
 \frac{2\lambda + 1}{\kappa^{(\lambda)}_0} \!=\!\sum_{j_1j_2}
 (f^{(\lambda)}_{\!j_1j_2})^2\!\left[
 \frac{(u^{(+)}_{j_1j_2})^2\varepsilon_{\!j_1j_2}^{(+)}(1\!-y^2_{\!j_1}\!-y^2_{\!j_2})}
       {(\varepsilon_{\!j_1j_2}^{(+)})^2-\omega^2}\!-\!
 \frac{(v^{(-)}_{\!j_1j_2})^2\varepsilon_{j_1j_2}^{(-)}(y^2_{\!j_1}\!-y^2_{\!j_2})}
       {(\varepsilon_{\!j_1j_2}^{(-)})^2-\omega^2}\right],
\end{gather}
where $f^{(\lambda)}_{\!j_1j_2}$ is the reduced single-particle
matrix element of the multipole operator $M_{\lambda \mu}$;
$\varepsilon^{(\pm)}_{j_1j_2}=\varepsilon_{j_1}\pm\varepsilon_{j_2}$,
 $u^{(+)}_{j_1j_2}=u_{j_1}v_{j_2} + v_{j_1}u_{j_2}$,
 $v^{(-)}_{j_1j_2}=u_{j_1}u_{j_2} - v_{j_1}v_{j_2}$.

Although at the present stage phonon amplitudes cannot be determined
unambiguously, the TQRPA Hamiltonian is diagonal in terms of thermal
phonon operators
\begin{equation}\label{diagonal}
\mathcal H_{\mathrm{TQRPA}} =\sum_{\lambda \mu i}\omega_{\lambda
i}(Q^\dag_{\lambda\mu i}Q^{\phantom{\dag}}_{\lambda\mu i}-
   \widetilde{Q}^\dag_{\lambda\mu i}\widetilde{Q}^{\phantom{\dag}}_{\lambda\mu
   i}).
\end{equation}
One can see that $\mathcal{H}_{\rm TQRPA}$ is invariant under the
following thermal Bogoliubov  transformation:
\begin{equation}\label{XY}
    Q_{\lambda\mu i}^\dag \to
    X_{\lambda i}Q_{\lambda\mu i}^\dag-Y_{\lambda i}\widetilde{Q}_{\lambda\mu i}^{\phantom{\dag}},
\qquad
    \widetilde{Q}_{\lambda\mu i}^\dag \to  X_{\lambda i}\widetilde{Q}_{\lambda\mu i}^\dag-
    Y_{\lambda i}Q_{\lambda\mu i}^{\phantom{\dag}}
\end{equation}
with $X^2_{\lambda i}-Y^2_{\lambda i}=1$. To fix the coefficients
$X_{\lambda i},~Y_{\lambda i}$ and finally determine the phonon
amplitudes $\psi, \widetilde{\psi}, \phi, \widetilde{\phi}, \eta,
\xi$, we again demand that the thermal phonon vacuum obey the
thermal state condition\footnote{Earlier, in \cite{IJMPE09} we have
used the other procedure to this aim. That procedure seems to be
much less evident and more lengthy.}. For $A$ in~(\ref{TSC}),  it is
convenient to take the multipole operator $M_{\lambda\mu}$. Then the
thermal state condition takes the form
\begin{equation}\label{TSC_QRPA}
M_{\lambda\mu}|0(T);\mathrm{ph}\rangle =
 {\rm e}^{\mathcal{ H}_{\mathrm{TQRPA}}/2T}\,
\widetilde{M}_{\lambda\mu}^\dag|0(T);\mathrm{ph}\rangle.
\end{equation}
Expressing $M_{\lambda\mu}$ through  phonon operators  we find the
coefficients $X_{\lambda i},~Y_{\lambda i}$
\begin{equation*}
  Y_{\lambda i}=\left[\exp\left(\frac{\omega_{\lambda
  i}}{T}\right)-1\right]^{-1/2},\qquad
  X_{\lambda i}=\bigl[1+Y_{\lambda i}^2\bigr]^{1/2}.
\end{equation*}
The coefficients $Y^2_{\lambda i}$ appear to be the thermal
occupation factors of the Bose-Einstein statistics. Thus, the phonon
amplitudes are dependent on both the types of thermal occupation
numbers: quasiparticle ones (the Fermi-Dirac type) and phonon ones
(the Bose-Einstein type). The expressions for all the phonon
amplitudes $\psi, \widetilde{\psi}, \phi, \widetilde{\phi}, \eta,
\xi$ can be found in \cite{IJMPE09}.

Once the structure of thermal phonons is determined, one can find
the $E\lambda$-transition strengths from the TQRPA thermal vacuum to
one-phonon states. The transition strengths to non-tilde and tilde
one-phonon states are related by
\begin{equation}\label{symmetry}
\widetilde\Phi^2_{\lambda i}=\exp(-\omega_{\lambda
i}/T)\Phi^2_{\lambda i}.
\end{equation}
This relation is equivalent to the principle of detailed balancing
connecting the probabilities for a probe to transfer energy $\omega$
to a heated system and to absorb energy $\omega$ from a heated
system.


Now we are ready to go beyond TQRPA and consider the effects of the
term $\mathcal H_{\rm qph}$ which is a thermal analogue of the
quasiparticle-phonon interaction \cite{Soloviev92}. It reads
\vspace{-3mm}
\begin{equation}\label{th_Ham3}
 {\mathcal H}_{\rm qph}=-
 \frac12\sum_{\lambda\mu i}\sum_{j_1j_2}
 \frac{f^{(\lambda)}_{j_1j_2}}{\sqrt{{\cal N}^{\lambda i}}}\left\{
 \bigl(Q^\dag_{\overline{\lambda\mu}i}\!+\!Q^{\phantom{\dag}}_{\lambda\mu i}\bigr)
  B_{\lambda\mu i}(j_1j_2)\!+\!({\rm h.c.})\!-\!({\rm t.c.})\right\}
\end{equation}
where the notation ''(h.c.)" and ''(t.c.)" stands for the items
which are hermitian-  and tilde-conjugate to the displayed ones;
${\cal N}^{\lambda i}$ is the normalization factor in the phonon
amplitudes. The operator $B_{\lambda\mu i}(j_1j_2)$ reads
\begin{multline*}
  B_{\lambda\mu i}(j_1j_2)=iu^{(+)}_{j_1j_2}\left(
  {\cal Z}_{j_1j_2}^{\lambda i}[\beta^\dag_{j_1}\widetilde{\beta}^{\phantom{\dag}}_{j_2}]^\lambda_\mu+
  {\cal Z}_{j_2j_1}^{\lambda i}[\widetilde{\beta}^\dag_{\overline{\jmath_1}}
                         \beta^{\phantom{\dag}}_{\overline{\jmath_2}}]^\lambda_\mu\right)-
  \\
  v^{(-)}_{j_1j_2}\left(
  {\cal X}_{j_1j_2}^{\lambda i}[\beta^\dag_{j_1}\beta^{\phantom{\dag}}_{\overline{\jmath_2}}]^\lambda_\mu+
  {\cal Y}_{j_1j_2}^{\lambda i}[\widetilde{\beta}^\dag_{\overline{\jmath_1}}
                         \widetilde{\beta}^{\phantom{\dag}}_{j_2}]^{\lambda}_{\mu}\right),
\end{multline*}
where the coefficients ${\cal X}_{j_1j_2}^{\lambda i}$, ${\cal
Y}_{j_1j_2}^{\lambda i}$ and ${\cal Z}_{j_1j_2}^{\lambda i}$ are the
following:
\begin{eqnarray*}
 \binom{\cal X}{\cal Y}_{j_1j_2}^{\lambda i}=x_{j_1}x_{j_2}\binom{X}{Y}_{\lambda i}+
                                   y_{j_1}y_{j_2}\binom{Y}{X}_{\lambda i}~,~~~
 {\cal Z}_{j_1j_2}^{\lambda i}=x_{j_1}y_{j_2}X_{\lambda i}+y_{j_1}x_{j_2}Y_{\lambda i}~.
 \end{eqnarray*}

The term $\mathcal H_{\rm qph}$ couples states with a different
number of thermal phonons. To take into account the phonon coupling,
we consider a trial wave function of the following form:
\begin{multline}\label{trial}
 |\Psi_\nu(JM)\rangle=\biggl[\sum_i\left\{ R_i(J\nu)\, Q^{\dag}_{JMi}+
 \widetilde R_i(J\nu)\, \widetilde Q^{\dag}_{\overline{JM}i}\right\}+
 \sum_{\stackrel{\lambda_1i_1}{\lambda_2i_2}}\Bigl\{P^{\lambda_1i_1}_{\lambda_2i_2}(J\nu)
 \bigl[Q^\dag_{\lambda_1i_1}Q^\dag_{\lambda_2i_2}\bigr]^J_M\\
 +\sum_{\stackrel{\lambda_1i_1}{\lambda_2 i_2}}S^{\lambda_1i_1}_{\lambda_2i_2}(J\nu)
 \bigl[Q^\dag_{\lambda_1i_1}\!\widetilde{Q}^\dag_{\overline{\lambda_2}i_2}\bigr]^J_M+
 \sum_{\stackrel{\lambda_1i_1}{\lambda_2 i_2}}\widetilde P^{\lambda_1i_1}_{\lambda_2i_2}(J\nu)
 \bigl[\widetilde{Q}^\dag_{\overline{\lambda_1}i_1}\widetilde{Q}^\dag_{\overline{\lambda_2}i_2}\bigr]^J_M\Bigr\}
 \biggr]|0(T);\mathrm{ph}\rangle.
\end{multline}
It should be stressed that in (\ref{trial}) we keep the thermal
vacuum of TQRPA. It means that we do not consider the influence of
phonon coupling on thermal occupation numbers. Note also that the
function (\ref{trial}) contains not only non-tilde one-phonon
components but the tilde ones as well. This is a new point in
comparison with Ref.~\cite{IJMPE09}.

The function (\ref{trial}) has to be normalized. This demand imposes
the following constraint on the amplitudes $R,\widetilde
R,~P,~S,~\widetilde P$:
\begin{multline}\label{constr2}
 \sum_i\left\{\bigl[R_i(J\nu)\bigr]^2+\bigl[\widetilde R_i(J\nu)\bigr]^2\right\}\\+
 \sum_{\stackrel{\lambda_1i_1}{\lambda_2i_2}}\left\{2\bigl[P^{\lambda_1i_1}_{\lambda_2i_2}(J\nu)\bigr]^2+
  \bigl[S^{\lambda_1i_1}_{\lambda_2i_2}(J\nu)\bigr]^2+
  2\bigl[\widetilde P^{\lambda_1i_1}_{\lambda_2i_2}(J\nu)\bigr]^2\right\}=1.
\end{multline}

Since the trial function contains three different types of
two-phonon components, there are three types of interaction matrix
elements which couple a thermal one-phonon state with two-phonon
ones
\begin{align}
  U^{\lambda_1i_1}_{\lambda_2i_2}(Ji)&=
  \langle 0(T);\mathrm{ph}|Q_{JMi}{\cal H}_{\rm qph}
  \bigl[Q^\dag_{\lambda_1i_1}Q^\dag_{\lambda_2i_2}\bigr]^J_M|0(T);\mathrm{ph}\rangle,
  \nonumber\\& \nonumber\\
  V^{\lambda_1i_1}_{\lambda_2i_2}(Ji)&=
  \langle 0(T);\mathrm{ph}|Q_{JMi}{\cal H}_{\rm qph}
  \bigl[Q^\dag_{\lambda_1i_1}
  \widetilde{Q}^\dag_{\overline{\lambda_2}i_2}\bigr]^J_M|0(T);\mathrm{ph}\rangle,\nonumber
  \\&\nonumber \\
  W^{\lambda_1i_1}_{\lambda_2i_2}(Ji)&=
  \langle 0(T);\mathrm{ph}|Q_{JMi}{\cal H}_{\rm qph}
  \bigl[\widetilde Q^\dag_{\overline{\lambda_1}i_1}
        \widetilde{Q}^\dag_{\overline{\lambda_2}i_2}\bigr]^J_M|0(T);\mathrm{ph}\rangle.
\end{align}
The expressions for the matrix elements
$U^{\lambda_1i_1}_{\lambda_2i_2}(Ji),
V^{\lambda_1i_1}_{\lambda_2i_2}(Ji)$, and
$W^{\lambda_1i_1}_{\lambda_2i_2}(Ji)$ via the phonon amplitudes
$\psi, \widetilde \psi, \phi$ etc. can be found in \cite{IJMPE09}.

Applying the variational principle to the average value of the
thermal Hamiltonian $\mathcal{H}_{\rm TQRPA}+ \mathcal{H}_{\rm qph}$
with respect to  $|\Psi_\nu(JM)\rangle$ under the normalization
constraint (\ref{constr2}) one gets a system of linear equations for
the amplitudes $R,\widetilde R,~P,~S,~\widetilde P$. The system has
a nontrivial solution if the energy $\eta_\nu$ of the state
$|\Psi_\nu(JM)\rangle$ obeys the following secular equation:
\begin{equation}\label{det}
\mathrm{det}\left|\begin{array}{cc}
  A(\eta_\nu) & B (\eta_\nu)\\
  B(-\eta_\nu) & A(-\eta_\nu)
\end{array}\right|=0,
\end{equation}
where
\begin{multline}
A_{ii'}(\eta_\nu)=
\bigl(\omega_{Ji}-\eta_\nu\bigr)\delta_{ii'}-\frac12\sum_{\stackrel{\lambda_1i_1}{\lambda_2i_2}}
 \left\{
 \frac{U^{\lambda_1i_1}_{\lambda_2i_2}(Ji)U^{\lambda_1i_1}_{\lambda_2i_2}(Ji')}
 {\omega_{\lambda_1i_1}+\omega_{\lambda_2i_2}-\eta_\nu} \right.\\+\left.
 2\frac{V^{\lambda_1i_1}_{\lambda_2i_2}(Ji)V^{\lambda_1i_1}_{\lambda_2i_2}(Ji')}
 {\omega_{\lambda_1i_1}-\omega_{\lambda_2i_2}-\eta_\nu}-
  \frac{W^{\lambda_1i_1}_{\lambda_2i_2}(Ji)W^{\lambda_1i_1}_{\lambda_2i_2}(Ji')}
  {\omega_{\lambda_1i_1}+\omega_{\lambda_2i_2}+\eta_\nu}\right\}
\end{multline}
and
\begin{multline}
B_{ii'}(\eta_\nu)=
\frac12\sum_{\stackrel{\lambda_1i_1}{\lambda_2i_2}}
 \left\{
 \frac{U^{\lambda_1i_1}_{\lambda_2i_2}(Ji)W^{\lambda_1i_1}_{\lambda_2i_2}(Ji')}
 {\omega_{\lambda_1i_1}+\omega_{\lambda_2i_2}-\eta_\nu} \right.\\+\left.
 2(-1)^{\lambda_1+\lambda_2+J}\frac{V^{\lambda_1i_1}_{\lambda_2i_2}(Ji)V^{\lambda_2i_2}_{\lambda_1i_1}(Ji')}
 {\omega_{\lambda_1i_1}-\omega_{\lambda_2i_2}-\eta_\nu}-
  \frac{W^{\lambda_1i_1}_{\lambda_2i_2}(Ji)U^{\lambda_1i_1}_{\lambda_2i_2}(Ji')}
  {\omega_{\lambda_1i_1}+\omega_{\lambda_2i_2}+\eta_\nu}\right\}.
\end{multline}

Physical effects which can be treated with the function
$|\Psi_\nu(JM)\rangle$ and Eq. (\ref{det}) relate to fragmentation
of basic nuclear excitations like quasiparticles and phonons, their
spreading widths and/or more consistent description of transition
strength distributions over a nuclear spectrum in hot nuclei.

The authors are thankful to Dr. V. Ponomarev for valuable
discussions and comments.


\end{document}